\documentclass[prd,aps,twocolumn,nofootinbib,preprintnumbers,superscriptaddress,preprintnumbers,balancelastpage,longbibliography]{revtex4-2}

\usepackage{graphicx}
\usepackage{bm}
\usepackage{hyperref}
\usepackage{hyperref}
\usepackage{slashed}
\usepackage{aas_macros}
\usepackage{float}
\usepackage{slashed}
\usepackage{lipsum}
\usepackage{subfigure}
\usepackage{multirow}
\usepackage{amsmath}
\usepackage{array} % for defining a new column type
\usepackage{varwidth} %for the varwidth minipage environment
\usepackage{tabularx}
\usepackage{braket}
\usepackage[dvipsnames]{xcolor}
\usepackage{wasysym}
\usepackage{fontawesome5}

\usepackage[normalem]{ulem}

\newcommand{\be}{\begin{equation}}
\newcommand{\ee}{\end{equation}}
\newcommand{\bea}{\begin{eqnarray}}
\newcommand{\eea}{\end{eqnarray}}

\hypersetup{
     colorlinks   = true,
     citecolor    = Orange,
     urlcolor     = Orange,
     linkcolor    = Orange
}

\usepackage{listings}
\usepackage{color,xcolor}

\begin{document}

%\preprint{UCI-HEP-TR-2024-02}

%\title{Spin-dependent Cross Section Constraints With Evaporation Effect in the Sun}
%\title{Spin-Dependent Constraints on Dark Matter Nuclear Scattering in the Sun Near the Evaporation Limit}
%\title{A Critical Analysis of Spin-Dependent Dark Matter Scattering \\in the Sun Near the Evaporation Limit}
\title{The Sun Can Strongly Constrain Spin-Dependent \\ Dark Matter Nucleon Scattering Below the Evaporation Limit}
%\title{Jupiter Metallic Hydrogen dark matter capture}
% another suggestion: Jovian Metallic Hydrogen as a Sub-GeV Dark Matter Trap

\author{Thong T.Q. Nguyen}
\thanks{{\scriptsize Email}: \href{mailto:thong.nguyen@fysik.su.se}{thong.nguyen@fysik.su.se}; \href{https://orcid.org/0000-0002-8460-0219}{0000-0002-8460-0219}}
\affiliation{Stockholm University and The Oskar Klein Centre for Cosmoparticle Physics, Alba Nova, 10691 Stockholm, Sweden}

\author{Tim Linden}
\thanks{{\scriptsize Email}: \href{mailto:linden@fysik.su.se}{linden@fysik.su.se};  \href{http://orcid.org/0000-0001-9888-0971}{0000-0001-9888-0971}}
\affiliation{Stockholm University and The Oskar Klein Centre for Cosmoparticle Physics, Alba Nova, 10691 Stockholm, Sweden}
\affiliation{Erlangen Centre for Astroparticle Physics (ECAP), Friedrich-Alexander-Universität \\ Erlangen-Nürnberg, Nikolaus-Fiebiger-Str. 2,
91058 Erlangen, Germany}

\begin{abstract}
\noindent The Sun is a promising target for dark matter (DM) searches due to its ability to accumulate DM particles via scattering and catalyze their self-annihilation. However, at low DM masses, DM particles can also ``evaporate" due to subsequent collisions with the hot thermal plasma of the Sun. While several modeling studies have calculated the competitive dynamics of DM evaporation and annihilation, observational studies have typically assumed a fixed 4~GeV ``evaporation limit", below which DM evaporates before it can annihilate. In this paper, we consider the competitive effects of DM evaporation and annihilation on spin-dependent DM nucleon cross-section limits, finding that Solar observations can continue to exceed terrestrial constraints by between 1--5 orders of magnitude for DM masses between 2--4~GeV, and can even provide world leading constraints below 0.2~GeV where direct detection is limited.
\end{abstract}

\maketitle

\section{Introduction}
\label{sect:intro}

Due to the overwhelming evidence that dark matter (DM) constitutes most of the matter in the Universe, identifying its non-gravitational interactions with Standard Model (SM) particles has become a key goal of the physics community~\cite{Bertone:2016nfn, Bertone:2004pz, Bertone:2018krk}. Notably, the Weakly Interacting Massive Particle (WIMP) still stands as one of the most promising candidates. While there are ongoing, terrestrial efforts to probe rare WIMP scattering interactions with the SM~\cite{XENON:2024wpa, LZ:2019sgr, PANDA-X:2024dlo, PICO:2019vsc}, early work by Press and Spergel~\cite{Press:1985ug} proposed an alternative and conceptually elegant idea: using the Sun as a ``natural'' detector of WIMP scattering. Specifically, WIMPs can scatter with nucleons and electrons inside the Sun, losing their kinetic energy and becoming gravitationally trapped. The accumulated WIMPs can subsequently annihilate and produce either neutrinos that escape the Sun, or long-lived mediators that decay into SM particles outside the Sun, yielding observable electromagnetic signals~\cite{Leane:2024bvh, Niblaeus:2019gjk}. 

Along these lines, physicists have used the impressive sensitivity of Water Cherenkov neutrino detectors like Super-Kamiokande (Super-K) and space-based $\gamma$-ray detectors like the Fermi-LAT, to probe the DM-nucleon scattering cross section, with sensitivities that, in many cases, surpass terrestrial DM searches~\cite{Kappl:2011kz, Bernal:2012qh, IceCube:2025fcu, Hooper:2025ohk, Berlin:2024lwe, ANTARES:2016obx, Widmark:2017yvd, Catena:2016ckl, IceCube:2016yoy, Acevedo:2020gro, Ng:1986qt, Krishna:2025ncv, Maity:2023rez, IceCube:2021xzo, Bell:2021esh, Bell:2011sn, Bell:2012dk, Kopp:2009et, Super-Kamiokande:2015xms} and $\gamma$-rays~\cite{Leane:2017vag, HAWC:2018szf, HAWC:2022khj, Nisa:2019mpb, HAWC:2018szf, Bose:2021cou, Bell:2021pyy, Serini:2022aed} and electrons~\cite{FermiLAT:2011ozd, Cuoco:2019mlb}. However, these studies often assume a hard cutoff in sensitivity for DM masses below 4~GeV~\cite{IceCube:2025fcu, Hooper:2025ohk, Berlin:2024lwe, ANTARES:2016obx, Widmark:2017yvd, Catena:2016ckl, IceCube:2016yoy, Acevedo:2020gro, Ng:1986qt, Krishna:2025ncv, Maity:2023rez, IceCube:2021xzo, Bell:2021esh, Bell:2011sn, Bell:2012dk, Kopp:2009et, Super-Kamiokande:2015xms, Leane:2017vag, HAWC:2018szf, HAWC:2022khj, Nisa:2019mpb, HAWC:2018szf, Bose:2021cou, Bell:2021pyy, Serini:2022aed, FermiLAT:2011ozd, Cuoco:2019mlb}. At smaller masses, captured DM particles can be ``evaporated" from the Sun by subsequent collisions with energetic Solar particles, decreasing the DM density and annihilation signal.

\begin{figure}[t!]
\centering
\includegraphics[width=1\columnwidth]{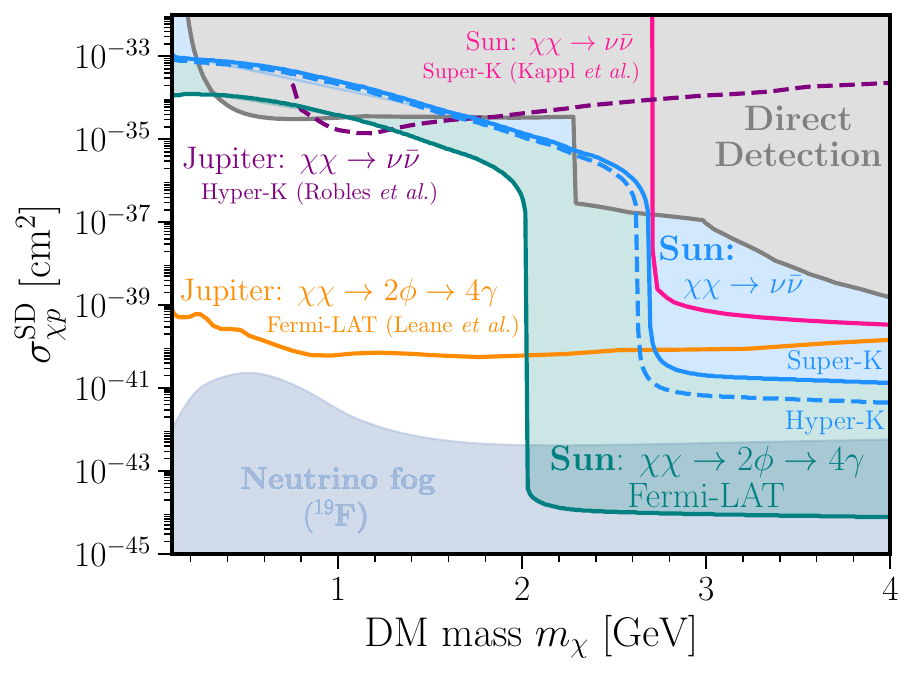}
\vspace{-0.8cm}
\caption{Spin-dependent cross section constraints (solid) and projections (dashed) for the Solar annihilation of DM particles with masses below 4~GeV to both neutrino and $\gamma$-ray final states. For the $\nu\bar{\nu}$ channel: Our Super-K constraints and Hyper-K projections are in blue, while a previous Super-K Solar analysis is in pink~\cite{Kappl:2011kz}, and the previous Hyper-K projection for Jupiter observations is in purple~\cite{Robles:2024tdh}. For $\gamma$-rays from the decay of long-lived mediators: Our Fermi-LAT constraint is in teal, while previous studies of Jupiter are shown in orange~\cite{Leane:2021tjj}. The combined limits from direct detection are in gray~\cite{PICO:2019vsc, PICASSO:2012ngj, CRESST:2022dtl, NEWS-G:2024jms}, while the neutrino fog is in cyan~\cite{OHare:2021utq}.}
\label{fig:constraint}
\vspace{-0.5cm}
\end{figure}

However, DM evaporation is a dynamic process that depends on the scattering cross-section as well as its velocity- and spin-dependence. While a 4~GeV threshold has been widely adopted by the observational community, detailed studies of DM evaporation find that the evaporation timescale for spin-dependent DM scattering typically falls below Jupiter's age near 3~GeV, decreasing precipitously from $\sim10^{19}$~s at 4~GeV to 10$^{10}$~s at 1~GeV. Using early Super-K data, several studies calculated the evaporation mass as a function of the scattering cross-section to explore DM annihilation below 4~GeV~\cite{Kappl:2011kz, Bernal:2012qh}. However, this still leaves a region where the evaporation timescale may fall below Jupiter's age, but the processes of DM evaporation and annihilation still occur on competitive timescales~\cite{Gould:1987ju, Garani:2017jcj, Kouvaris:2015nsa, Busoni:2017mhe, Garani:2017jcj}, which can be investigated with current neutrino and $\gamma$-ray data.

In this paper, we use existing Super-K data to produce a self-consistent analysis of current neutrino and $\gamma$-ray constraints on solar DM capture and annihilation in the regime below the typically adopted 4~GeV solar evaporation cutoff. Focusing on the spin-dependent nuclear scattering cross section and assuming an $s$-wave, thermally averaged, DM annihilation cross-section, we derive world leading constraints and projections on DM scattering for DM masses between 2--4~GeV, which exceed terrestrial constraints by more than an order of magnitude. Despite weakening considerably at lower masses, our constraints again provide world leading sensitivity for DM masses below 0.2~GeV, where terrestrial constraints are extremely weak. We show our results in Fig.~\ref{fig:constraint}, finding that our constraints can, in some cases, extend all the way into the neutrino-fog, where terrestrial direct detection strategies are significantly hampered~\cite{OHare:2021utq, Chou:2022luk}.

This paper is organized as follows. In Section~\ref{sect:CaptEva}, we present our calculations of DM capture and evaporation. In Section~\ref{sect:AnnRate}, we discuss the regime where the equilibrium assumption breaks down and compute the resulting annihilation rate. In Section~\ref{sect:signal}, we calculate benchmark DM annihilation signals from the Sun and compare them with current neutrino and $\gamma$-ray observations. In Section~\ref{sect:result}, we derive constraints and projections for the spin-dependent DM–proton cross section. We conclude in Section~\ref{sect:conclusion}.

\begin{figure*}[tbp!]
\centering
\includegraphics[width=2\columnwidth]{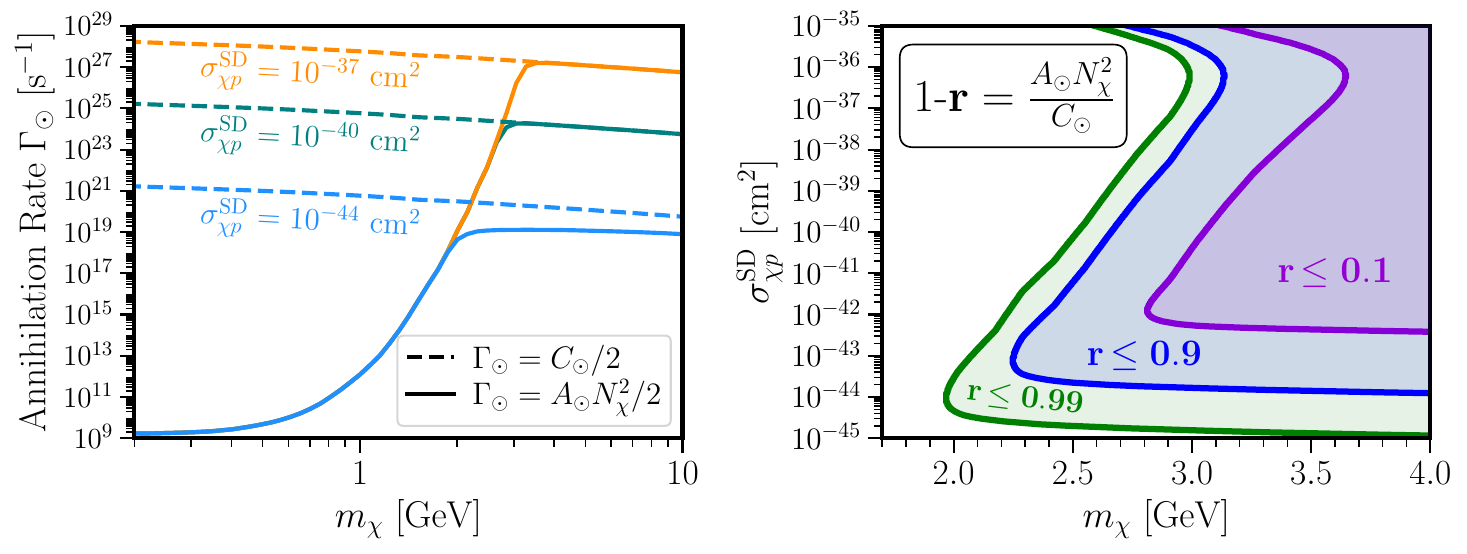}
\vspace{-0.5cm}
\caption{(\textbf{Left}) Total DM annihilation rates (solid) for spin-dependent cross sections of $10^{-37}$~cm$^{2}$ (orange), $10^{-40}$~cm$^{2}$ (teal), and $10^{-44}$~cm$^{2}$ (light blue). Dashed lines show the equilibrium approximation. (\textbf{Right}) Spin-dependent cross section regions where the true annihilation rate is reduced by no more than 10\% (violet), 90\% (blue) or 99\% (green) from half the DM-capture rate, which corresponds to the assumed annihilation rate under the equilibrium assumption.} 
\label{fig:AnnRate}
\vspace{-0.5cm}
\end{figure*}

\section{DM Capture and Evaporation in the Sun with Spin-Dependent Cross Sections}
\label{sect:CaptEva}

We consider DM scattering with protons inside the Sun through spin-dependent interactions~\cite{Gori:2025jzu, Hooper:2025iii, Giffin:2025hdx, Carena:2004xs, Blanco:2019hah, Beenakker:2025mhf, DiFranzo:2013vra, Lin:2019uvt}, which governs both the capture and evaporation of DM in the solar interior. For a contact interaction mediated by an axial current, the total DM–nucleus cross section at zero momentum transfer, and for a nucleus with mass number $A$ and total angular momentum $J_{A}$, is given by
\begin{equation}
\sigma^{\rm SD}_{\chi A }=\left( \frac{\tilde{\mu}_{\chi A}}{\tilde{\mu}_{p}}\right)^{2}\frac{4 (J_{A}+1)}{3J_{A}}|\langle S_{p A}\rangle +\langle S_{n A}\rangle|^{2}\sigma^{\rm SD}_{\chi p},
\label{eq:sig}
\end{equation}
where we assume that DM couples equally to protons and neutrons. The quantities $\tilde{\mu}{\chi A}$ and $\tilde{\mu}{p}$ denote the reduced masses of the DM–nucleus and DM–proton systems, respectively. We take the proton and neutron spin expectation values, averaged over all nucleons, $\langle S_{p A}\rangle$ and $\langle S_{n A}\rangle$, from Refs.~\cite{Engel:1989ix, Ellis:1987sh, Pacheco:1989jz, Engel:1992bf, Divari:2000dc, Bednyakov:2004xq}.

With this interaction, the rate for DM to change its velocity from $w$ to $v$ through a single scatter is
\begin{align}
    R^{\pm}(w\to v)&=\frac{2}{\sqrt{\pi}}\frac{\mu_{p,+}^{2}}{\mu_{p}}\frac{v}{w}n_{p}(r)\sigma_{\chi A}\label{eq:Rate}\\
    &\times \left[ \chi(\pm \alpha_{-}, \alpha_{+})+\chi(\pm \beta_{-},\beta_{+})e^{\mu_{p}(w^{2}-v^{2})/u_{p}^{2}} \right],\nonumber
\end{align}
where we assume an isotropic and velocity-independent DM cross section with the target nuclei, which are predominantly hydrogen. We define the mass ratios \mbox{$\mu_{p}=m_{\chi}/m_{H}\simeq m_{\chi}/m_{p}$} and $\mu_{p,\pm}=(\mu_{p}\pm 1)/2$. The ``$+$'' sign corresponds to $w<v$, for the DM capture process, while the ``$-$'' sign corresponds to $w>v$ for the DM evaporation process. We adopt the standard solar model from Ref.~\cite{Magg:2022rxb} for the proton number density $n_{p}(r)$ and the solar temperature profile $T_{\odot}(r)$. The most probable proton speed is $u_{p}(r)=\sqrt{2T_{\odot}(r)/m_{p}}$. The explicit forms of the functions $\chi$, $\alpha$, and $\beta$ are taken from Ref.~\cite{Garani:2017jcj}.

We calculate the capture rate of DM particles with mass $m_{\chi}$ that travel from the halo toward the Sun with velocity $u_{\chi}$, fall into the solar gravitational potential, scatter with protons, and become gravitationally bound. For weakly interacting DM, the capture rate is~\cite{Gould:1987ju, Garani:2017jcj}
\begin{align}
    C_{\rm weak}&=\int_{0}^{R_{\odot}}{\rm d}r4\pi r^{2}\left(\frac{\rho_{\chi}^{\odot}}{m_{\chi}}\right)\int_{0}^{\infty}{\rm d}u_{\chi}\frac{f_{v_{\odot}}(u_{\chi})}{u_{\chi}}w(r)\nonumber\\
    &\times \int_{0}^{v_{e}(r)}{\rm d}vR^{-}(w\to v)|F_{A}(q)|^{2},
    \label{eq:Cweak}
\end{align}
where we integrate the final DM velocity $v$ over values below the local escape velocity $v_{e}(r)$. We adopt a local solar DM density of $\rho_{\chi}^{\odot}=0.3$~GeV/cm$^{3}$ and set the nuclear form factor for hydrogen targets to $F_{A}(q)=F_{H}=1$. The DM speed in the solar gravitational potential prior to scattering is
\begin{equation}
    w(r)=\sqrt{u_{\chi}^{2}+v_{e}^{2}(r)}.
    \label{eq:wfunct}
\end{equation}
For the capture process, we adopt the standard Maxwell–Boltzmann halo velocity distribution for incoming DM, $f_{v_{\odot}}$, from Ref.~\cite{Folsom:2025lly}, with an upper velocity cutoff given by the Galactic escape speed $u_{\chi}^{\rm gal}\simeq 600$~km/s~\cite{Piffl:2013mla}. We then rescale the capture rate by the geometric saturation limit, $C_{\odot}^{\rm geo}$, from Refs.~\cite{Garani:2017jcj, Nguyen:2025ygc, Bernal:2012qh} as
\begin{equation}
    C_{\odot}=C_{\rm weak}\left( 1 - e^{-C_{\odot}^{\rm geo}/C_{\rm weak}} \right),
    \label{eq:Ctotal}
\end{equation}
which accounts for the transition to the optically thick regime. Our result agrees with the calculation in Ref.~\cite{Garani:2017jcj}. We note that several public codes are available for computing DM capture in the Sun, including \texttt{Asteria}~\cite{Leane:2023woh}, \texttt{DarkCapPy}~\cite{Green:2018qwo}, \texttt{WimPyC}~\cite{Kang:2025yci}, and \texttt{DarkSUSY}~\cite{Bringmann:2018lay}.

After capture, the WIMP temperature depends on its scattering cross section with the proton targets. In the low cross section regime, WIMPs scatter throughout the Sun and maintain a spatially constant temperature, corresponding to the isothermal regime. We compute this temperature, $T_{\chi}=T_{\chi}^{\rm iso}$, following Refs.~\cite{Garani:2017jcj, Busoni:2017mhe, Blanco:2025wpo}. In the strong interaction regime, DM particles thermalize locally and enter the local thermal equilibrium (LTE) regime, with a temperature $T_{\chi}=T_{\chi}^{\rm LTE}=T_{\odot}(r)$. Using these temperature prescriptions, we calculate the DM spatial distribution $n_{\chi}(r)$ following Refs.~\cite{Garani:2017jcj, Busoni:2017mhe}. We interpolate between the isothermal and LTE regimes using the Knudsen number for each cross section value~\cite{Gould:1987ju, Gould:1989tu, Gould:1989ez, DeRocco:2022rze, Leane:2022hkk}. We self-consistently employ this DM distribution for both the evaporation and annihilation calculations of WIMP particles inside the Sun.

We calculate the total rate at which a captured DM particle with a velocity below the local escape velocity scatters with thermal protons, gains sufficient energy, and evaporates from the Sun. Following Ref.~\cite{Garani:2017jcj}, the evaporation rate is
\begin{align}
    E_{\odot}&=\int_{0}^{R_{\odot}}{\rm d}r\,s(r)n_{\chi}(r)4\pi r^{2}\int_{0}^{v_{e}(r)}{\rm d}w\, 4\pi w^{2} f_{\chi}(w, r)\nonumber\\
    &\times \int_{v_{e}(r)}^{\infty} {\rm d}v\, R^{+}(w\to v),
    \label{eq:Eva}
\end{align}
where we assume that the DM velocity distribution $f_{\chi}(w,r)$ follows a Maxwell–Boltzmann distribution. We compute the suppression factor $s(r)$ following Refs.~\cite{Garani:2017jcj, Busoni:2017mhe, Leane:2022hkk, Leane:2024bvh}. Our evaporation rate agrees with the calculation in Ref.~\cite{Garani:2017jcj}, including the latest erratum~\cite{Garani_2025}, for spin-dependent DM–proton scattering.

\section{The Competition Between Evaporation and Annihilation}
\label{sect:AnnRate}

In addition to upscattering by fast-moving protons, captured DM particles, with large population inside the Sun, can annihilate with each other. The annihilation rate for a pair of DM particles is
\begin{align}
    A_{\odot} = \langle \sigma v\rangle_{\chi\chi}\frac{\int_{0}^{R_{\odot}}{\rm d}r\, 4\pi r^{2} n_{\chi}^{2}(r)}{\left( \int_{0}^{R_{\odot}}{\rm d}r\, 4\pi r^{2} n_{\chi}(r) \right)^{2}},
    \label{eq:ARate}
\end{align}
where $\langle \sigma v\rangle_{\chi\chi}$ denotes the thermally averaged DM annihilation cross section. Following previous studies, we adopt the $s$-wave thermal annihilation cross section of $\langle \sigma v\rangle_{\chi\chi}=3\times10^{-26}$~cm$^{3}$/s.

We then compute the equilibrium timescale between DM capture and annihilation~\cite{Gaisser:1986ha, Griest:1986yu} as
\begin{equation}
    t_{\rm eq} = 1 /\sqrt{C_{\odot}A_{\odot}},
    \label{eq:teq}
\end{equation}
for which the annihilation rate equals half of the capture rate, $\Gamma_{\odot}=C_{\odot}/2$. However, for DM masses below 4~GeV, evaporation becomes important~\cite{Garani:2021feo}, and the annihilation rate depends on the total number of captured DM particles, $N_{\chi}(t)$. The resulting annihilation rate at the solar age is given by~\cite{Garani:2017jcj, Busoni:2017mhe}:
\begin{align}
    \Gamma_{\odot}&=\frac{1}{2}A_{\odot}N_{\chi}^{2}(t=t_{\odot})\label{eq:Gamma}\\
    &=\sqrt{\frac{C_{\odot}}{A_{\odot}}}\frac{\tanh (\kappa t_{\odot}/t_{\rm eq})}{\kappa + \frac{1}{2}E_{\odot}t_{\rm eq}\tanh(\kappa t_{\odot}/t_{\rm eq})},\nonumber
\end{align}
where $\kappa=\sqrt{1 + (E_{\odot}t_{\rm eq}/2)^{2}}$ and $t_{\odot}\simeq 4.57$~Gyrs denotes the age of the Sun.

We compute the total DM annihilation rate for three representative cross section benchmarks in Fig.~\ref{fig:AnnRate} (left). As in Ref.~\cite{Garani:2017jcj}, the equilibrium assumption can begin to fail in two scenarios: (1) when DM capture proceeds too slowly and the density for capture/annihilation equilibrium has not been reached by the current age of the object, and (2) when evaporation becomes an important competing force that removes DM particles before they can annihilate. We note three important rules of thumb. First, evaporation begins to be a significant issue at masses below around 3~GeV. Second, the exact evaporation value depends on the scattering cross-section, with lower scattering cross-sections leading to \emph{lower} evaporation masses through most of the cross-section range we are concerned with (though see Ref.~\cite{Leane:2022hkk}). Third, however, there is a scattering cross section below which equilibrium is never reached, with only a mild WIMP mass dependence. For the sun, this tends to occur near a spin-dependent cross section of $\sim10^{-44}$~cm$^{2}$. Below this value, solar WIMP searches are still possible, but the sensitivity will typically start decaying as $\left(\sigma^{\rm SD}_{\chi p}\right)^2$ rather than $\sigma^{\rm SD}_{\chi p}$.

In Fig.~\ref{fig:AnnRate} (right), we present the above information as contours in the DM–proton cross section, which represent regions where the ratio between the true WIMP annihilation rate and the prediction from the equilibrium approximation are larger or equal to 99\%, 10\%, and 1\%. We find that for cross sections below $10^{-42}$~cm$^{2}$, the equilibrium approximation begins to break down for DM masses above 3.2~GeV. These results demonstrate that, for very small cross sections (typically below $10^{-44}$~cm$^{2}$), annihilation rate calculations must treat non-equilibrium effects explicitly, as the equilibrium approximation no longer applies. Notably, for cross sections higher than $10^{-36}$~cm$^{2}$, DM scattering increasingly lies in the LTE regime. In this regime, higher cross-sections lead to smaller evaporation masses, due to the fact that multi-scattering causes DM particles that ``evaporate" from the core, to subsequently scatter and re-cool in the lower temperature atmosphere.\footnote{We thank Sergio Palomares-Ruiz for pointing out this regime.} We note that our results agree with previous studies from Refs.~\cite{Garani:2017jcj, Busoni:2017mhe, Leane:2022hkk}.

\begin{figure*}[tbp]
\centering
\includegraphics[width=1\columnwidth]{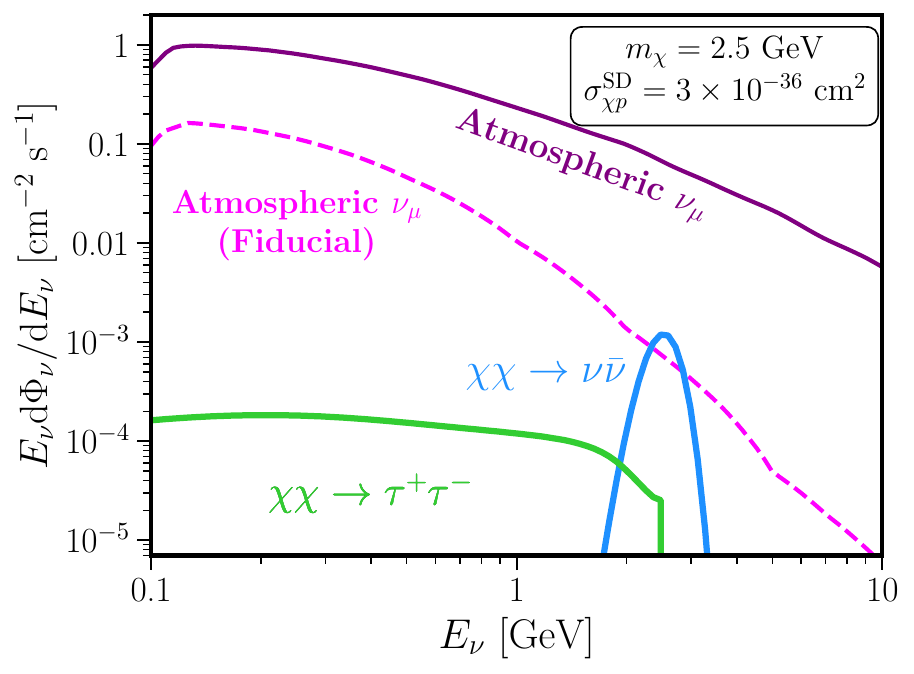}
\includegraphics[width=1\columnwidth]{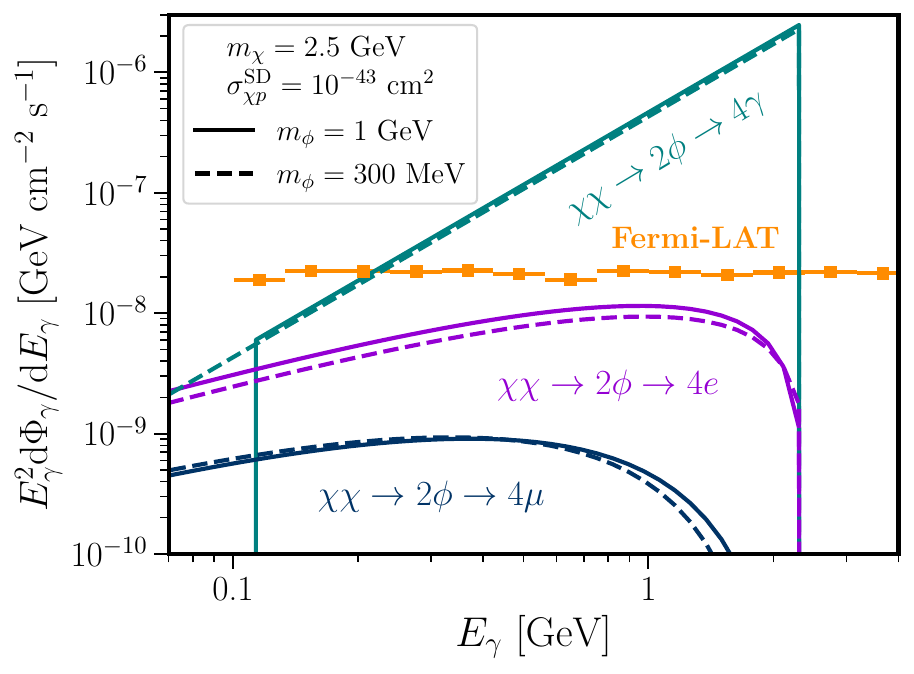}
\vspace{-0.5cm}
\caption{DM annihilation signals from the Sun for a DM mass $m_{\chi}=2.5$~GeV using neutrino observations ({\bf Left}), which include a diffuse atmospheric muon neutrino background (purple)~\cite{Super-Kamiokande:2015qek}, with angular cuts corresponding to Super-K's angular resolution (magenta)~\cite{Nguyen:2025ygc}. The neutrino fluxes from DM annihilation to $\nu\bar{\nu}$ (light blue) and $\tau^{+}\tau^{-}$ (lime) are shown for a scattering cross-section of $\sigma^{\rm SD}_{\chi p}=3\times10^{-36}$~cm$^{2}$. ({\bf Right:}) Results for $\gamma$-ray observations, with the Fermi-LAT solar disk $\gamma$-ray flux shown in orange. The $\gamma$-ray flux from DM annihilation to long-lived mediators $\phi\phi$ are shown for mediator decays into $\gamma\gamma$ (teal), $e^{+}e^{-}$ (violet) and $\mu^{+}\mu^{-}$ (dark blue) for a scattering cross-section of $\sigma^{\rm SD}_{\chi p}=10^{-43}$~cm$^{2}$. Results for $m_{\phi}=1$~GeV (solid lines) and for $m_{\phi}=300$~MeV (dashed lines) demonstrate the advantages of $\gamma$-ray observations for DM signals below the evaporation limit.}
\label{fig:observation}
\vspace{-0.5cm}
\end{figure*}

\section{Indirect Detection Signal}
\label{sect:signal}
Even though evaporation quickly becomes relevant (and then dominant) below 4~GeV, a substantial population of DM particles can remain trapped in the Sun on long time-scales. These captured DM particles can annihilate and produce $\gamma$-rays and neutrinos, with fluxes given by~\cite{Nguyen:2022zwb, Acevedo:2023xnu, Leane:2021ihh}
\begin{equation}
    \frac{{\rm d}\Phi_{i}}{{\rm d}E_{i}}=\frac{\Gamma_{\odot}}{4\pi D_{\odot}^{2}}\times\frac{{\rm d}N_{i}}{{\rm d}E_{i}}\times P_{\rm surv}^{i},
\end{equation}
where $i\equiv \nu,\gamma$ and $D_{\odot}=1$~AU is the distance from the Sun to the detector. The quantity ${\rm d}N_{i}/{\rm d}E_{i}$ denotes the energy spectrum of the annihilation products. We compute the survival probability $P_{\rm surv}^{i}$ based on whether the signal consists of neutrinos produced promptly inside the Sun or $\gamma$-rays originating from mediator decays outside the solar atmosphere.

\subsection{Neutrino Constraints}
\label{subsect:neutrino }
WIMP annihilation in the Sun can produce neutrinos either directly or through the secondary decays of SM final states. For DM masses below 4~GeV, we focus on two dominant annihilation channels, $\nu\bar{\nu}$ and $\tau^{+}\tau^{-}$. DM can also annihilate into $e^{+}e^{-}$, $\mu^{+}\mu^{-}$, and $\pi^{+}\pi^{-}$ final states, which generate neutrinos via weak bremsstrahlung or particle decays. We do not include these channels in our analysis. Notably, within the dense solar environment, muons and pions produced via DM decay can cool significantly before the muons or pions decay, shifting the resulting neutrino spectrum below the 100~MeV energy threshold adopted in our analysis~\cite{Rott:2012qb, Bernal:2012qh}.

For the $\nu\bar{\nu}$ channel, the neutrino spectrum is a Dirac delta function and is smeared primarily by the energy resolution of the Super-K and Hyper-K detectors, which we assume to be Gaussian distributed with a standard deviation of 10\% of the neutrino energy~\cite{Nguyen:2025ygc}. For the $\tau^{+}\tau^{-}$ channel, we compute the neutrino spectrum using the \texttt{PPPC4DM$\nu$} package~\cite{Baratella:2013fya}. We note that other public tools, such as \texttt{$\chi$aro$\nu$}~\cite{Liu:2020ckq} and \texttt{DarkSUSY}~\cite{Bringmann:2018lay}, yield consistent results. We include neutrino oscillation effects by averaging the total spectrum with a factor of $1/3$ to obtain the muon neutrino spectrum.

Neutrinos produced by DM annihilation inside the Sun can scatter with hydrogen along their path to the solar surface. We use the neutrino–hydrogen cross section $\sigma_{\nu H}$ from Ref.~\cite{Zhou:2023mou} to compute the scattering rate and derive the neutrino survival probability~\cite{Nguyen:2025ygc},
\begin{equation}
    P_{\rm surv}^{\nu}(E_{\nu})=1 - e^{-R_{\odot}/\int_{0}^{R_{\odot}}{\rm d}r\,\sigma_{\chi H}(E_{\nu})\, n_{H}(r)},\label{eq:Psurv}
\end{equation}
where we take the number density of Hydrogen atoms to be \mbox{$n_{H}\simeq n_{p}$}. We find that the survival probability exceeds 99\% for neutrino energies below 4~GeV.

In Fig.~\ref{fig:observation} (left), we show the muon neutrino fluxes from the two DM annihilation channels considered, for a benchmark DM mass of $m_{\chi}=2.5$~GeV and a spin-dependent cross section of $\sigma^{\rm SD}_{\chi p}=3\times10^{-36}$~cm$^{2}$. We also show the all-sky diffuse atmospheric muon neutrino background in the energy range 0.1--10~GeV, which constitutes the dominant background for terrestrial neutrino detectors~\cite{Super-Kamiokande:2015qek}. For water Cherenkov detectors such as Super-K, we use the average angular resolution from Refs.~\cite{Konishi:2010mv, Galkin:2008qe} to reduce the background to an effective value that could be confused with true solar emission, which we refer to as the ``fiducial" flux. Notably, at this benchmark mass where the total DM annihilation rate is suppressed due to evaporation, the Sun can still provide detectable neutrino signal for water Cherenkov detectors.

Using both the muon neutrino flux from DM annihilation and the fiducial diffuse background, we compute the expected number of observed neutrino events in water Cherenkov detectors in the energy range $[E_{\nu}^{\rm min}, E_{\nu}^{\rm max}]$ as
\begin{equation}
    N_{\nu}^{\chi/{\rm bkg}}=\int_{E_{\nu}^{\rm min}}^{E_{\nu}^{\rm max}}{\rm d}E_{\nu}\frac{{\rm d}\Phi_{\nu}^{\chi/{\rm bkg}}}{{\rm d}E_{\nu}}\sigma_{\nu H_{2}0}(E_{\nu})\times \xi,
\end{equation}
where we use the neutrino–water cross section from Ref.~\cite{Zhou:2023mou}. The detector exposure is $\xi=N_{H_{2}O}\times T_{\rm obs}$, where $N_{H_{2}O}$ is the number of water molecules in the fiducial volume and $T_{\rm obs}$ is the observation time. We assume an exposure time of 10~years and fiducial masses of 22.5~kton of water for Super-K and 187~kton for Hyper-K.

We compute the expected number of neutrino events for both the (fiducial) background and the DM signal. We then derive 95\% CL limits on the DM–proton cross section by requiring the two event counts to be statistically distinguishable, using Poisson statistics~\cite{Nguyen:2025ygc},
\begin{equation}
    \sum_{k=N^{\chi}_{\nu}+N^{\rm bkg}_{\nu}}^{\infty}\frac{\lambda^{k}e^{-\lambda}}{\Gamma(k+1)}\leq 0.05,
\end{equation}
where $\lambda=N^{\rm bkg}_{\nu}$ denotes the expected number of background events. For the null detection of neutrinos from the Sun, we fix the background expectation to $\lambda=2.7$. We note that an additional, but highly subdominant, background arises from solar atmospheric neutrinos produced by high-energy cosmic rays interacting with the solar atmosphere~\cite{Edsjo:2017kjk, delaTorre:2025llo, Ng:2017aur, Mazziotta:2020uey}.

\subsection{Gamma-Ray Constraints}
\label{subsect:gamma}

Captured DM can also produce $\gamma$-rays through annihilations into long-lived mediators. Because these mediators interact only weakly with hydrogen inside the Sun, they can escape the solar interior and subsequently decay into SM particles~\cite{Leane:2017vag, Leane:2021ihh, Nguyen:2025tkl, Nisa:2019mpb, DelaTorreLuque:2025zjt, Chu:2017vao}. We consider the case in which the mediator is a scalar particle that decays into a pair of SM particles, such as photons, $e^{\pm}$, or $\mu^{\pm}$, producing $\gamma$-ray spectra in the GeV range.

For the $2\gamma$ decay channel, we adopt the box-shaped gamma-ray spectrum from Refs.~\cite{Leane:2021tjj, Abdullah:2014lla},
\begin{equation}
    \frac{{\rm d}N_{\gamma}}{{\rm d}E_{\gamma}}\Big{|}_{\rm \chi\chi\to2\phi \to 4\gamma}=\frac{4}{\Delta E}\Theta(E_{\rm max}-E)\Theta(E-E_{\rm min}),
    \label{eq:dnde_box}
\end{equation}
where $\Delta E=\sqrt{m_{\chi}^{2}-m_{\phi}^{2}}$. The photon number density is flat in the interval
\begin{equation}
    E_{\rm max/min}=\frac{m_{\chi}}{2}\left( 1 \pm \sqrt{1 - \frac{m_{\phi}^{2}}{m_{\chi}^{2}}} \right).
\end{equation}

For mediator decays into $e^{\pm}$ and $\mu^{\pm}$ final states, we first compute the photon spectra in the mediator rest frame. In the $e^{\pm}$ channel, photons are produced via final-state radiation, for which we use the expressions in Refs.~\cite{Bystritskiy:2005ib, Cirelli:2020bpc, Nguyen:2024kwy}. In the $\mu^{\pm}$ channel, photons are generated through both final-state radiation in the mediator rest frame and through muon decays into electrons, which produce an additional photon component in the muon rest frame. We calculate both contributions using the formulae in Ref.~\cite{Coogan:2019qpu}. We validate our results against the \texttt{Hazma} package~\cite{Coogan:2019qpu, Coogan:2021sjs}. Finally, we obtain the spectra in the laboratory frame by applying the transformation
\begin{equation}
    \frac{{\rm d N_{\gamma}}}{{\rm d}E_{\gamma}}\Big{|}_{\rm lab}=\int^{\theta_{\rm lab}^{\rm max}}_{1}{\rm d}c\theta_{\rm lab}\frac{1}{2\eta (\beta_{L}\,c\theta_{\rm lab}-1)}\frac{dN_{\gamma}}{{\rm d}E_{\gamma}}\Big{|}_{\rm rest},
\end{equation}
where $c\theta{\rm lab}\equiv\cos\theta_{\rm lab}$, $\eta\simeq m_{\chi}/m_{\phi}$ is the Lorentz boost factor, and $\beta_{L}=\sqrt{1-1/\eta^{2}}$. The photon energies in the two frames are related by
\begin{equation}
    E^{\rm rest}_{\gamma}=\eta E^{\rm lab}_{\gamma}(1 - \beta_{L}\, c\theta_{\rm lab}).
\end{equation}
We use this relation to determine the upper limit $c\theta_{\rm lab}^{\rm max}$ of the angular integration in our Lorentz boost.

To observe $\gamma$-rays from the Sun with space-based detectors such as the Fermi-LAT, boosted mediators must decay outside the solar radius but before reaching the Earth. This requirement our delay length satisfies
\begin{equation}
    R_{\odot}\leq L_{\phi}\simeq \eta c \tau_{\phi}\leq D_{\odot},\label{eq:DecayLength}
\end{equation}
where $\tau_{\phi}$ denotes the mediator lifetime. The corresponding survival probability for mediator decays into Standard Model particles is
\begin{equation}
    P_{\rm surv}^{\phi\to {\rm SM}} = e^{-R_{\odot}/L_{\phi}} - e^{-D_{\odot}/L_{\phi}}.
\end{equation}
This probability is approximately 0.4 for decays near the solar surface, rises to a maximum close to unity at a distance of $\sim 10,R_{\odot}$, and then decreases to about 0.6 near the Earth’s orbit~\cite{Leane:2017vag}. This probability is also the key term which demands that the decay lifetime is on the order of the Sun-Earth distance, as it falls to 0 for both decay lifetimes that are much larger than the Earth-Sun distance or much smaller than the radius of the Sun. In our analysis, we adopt the most optimistic case in which this probability is unity to facilitate comparison with previous studies.

In Fig.~\ref{fig:observation} (right), we present the $\gamma$-ray differential energy flux from DM annihilation in the Sun into long-lived mediators that subsequently decay into $2\gamma$, $e^{\pm}$, and $\mu^{\pm}$ final states. We adopt the same benchmark DM mass, $m_{\chi}=2.5$~GeV, but a smaller spin-dependent cross section of $\sigma^{\rm SD}_{\chi p}=10^{-43}$~cm$^{2}$. We compute the fluxes for two representative mediator masses, $m_{\phi}=300$~MeV and $1$~GeV, and find that the results are similar. We then compare our predictions with current solar $\gamma$-ray observations from the Fermi-LAT~\cite{Linden:2020lvz}, which provide data in the energy range 0.1--100~GeV with an energy-averaged upper bound on the $\gamma$-ray energy flux of approximately $2\times10^{-8}$~GeV~cm$^{-2}$~s$^{-1}$. We derive upper limits on the spin-dependent DM–proton cross section for each annihilation channel that yields a total gamma-ray energy flux exceeding the Fermi-LAT data. We note that this limit is quite conservative, as the majority of the solar $\gamma$-ray emission is thought to be astrophysical in nature~\cite{Puzzoni:2025hdk, Li:2025xxi, Dorner:2025ait}, stemming from the interactions of cosmic-ray protons with the solar surface. Notably, there are time- and angularly- dependent variations in the solar $\gamma$-ray flux that preclude a DM-dominated origin~\cite{Linden:2020lvz}.

\section{Constraints on Spin-Dependent DM-nucleon Cross section}
\label{sect:result}

In Fig.~\ref{fig:constraint}, we present constraints from Super-K and Fermi-LAT data, as well as near-future projections for Hyper-K. Our current results strongly constrain the spin-dependent WIMP–proton cross section in the DM mass range 0.1--4~GeV for annihilation channels that produce only photon and neutrino final states. We also show the current strongest limits from direct detection, obtained by combining results from PICO60~\cite{PICO:2019vsc}, PICASSO~\cite{PICASSO:2012ngj}, CRESST~\cite{CRESST:2022dtl}, and NEWS-G~\cite{NEWS-G:2024jms}. For comparison, we show the previous constraint for Super-K observations of the $\nu\bar{\nu}$ annihilation channel from the Sun~\cite{Kappl:2011kz}. We also include previous Fermi-LAT constraints~\cite{Leane:2021tjj} and Hyper-K projections~\cite{Robles:2024tdh} derived from Jupiter observations using the same annihilation channels.

\begin{figure}[tbp]
\centering
\includegraphics[width=1\columnwidth]{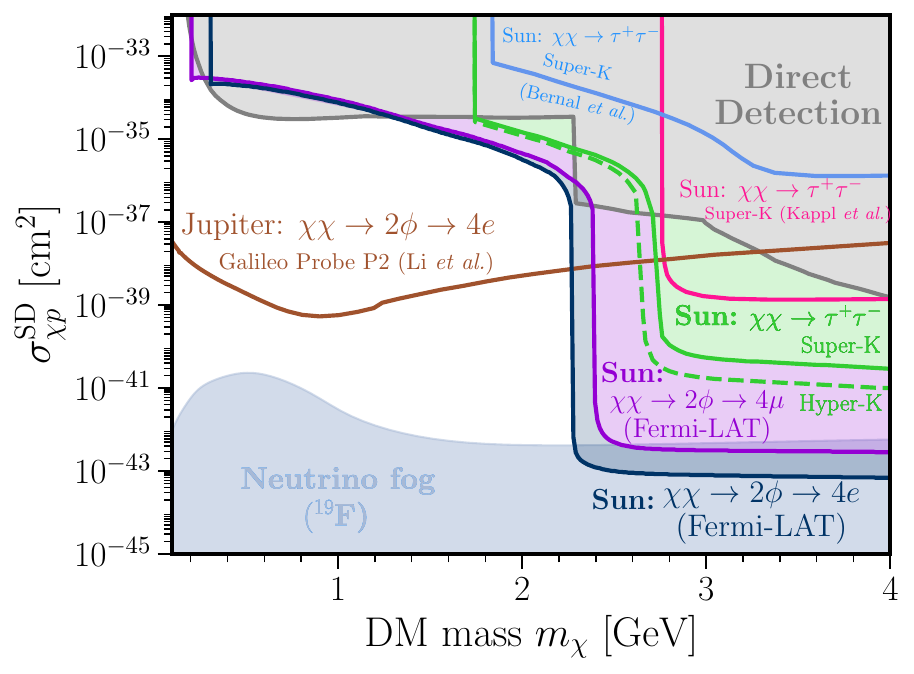}
\vspace{-0.5cm}
\caption{Similar to Fig.~\ref{fig:constraint}, but for leptonic final states. Our results using neutrinos observations of DM annihilation to the $\tau^{+}\tau^{-}$ channel are in lime, while the previous Super-K limits for the same channel are in pink~\cite{Kappl:2011kz} and light blue~\cite{Bernal:2012qh}. Our Fermi-LAT constraints for long-lived mediator decay to $e^{+}e^{-}$ are in navy blue, and $\mu^{+}\mu^{-}$ are in dark violet. Previous results for long-lived mediator decay to $e^{+}e^{-}$ outside Jupiter from Galileo Probe P2 observations are in brown~\cite{Li:2022wix, Yan:2023kdg, Singh:2024nou}. }
\label{fig:constraint_lepton}
\vspace{-0.5cm}
\end{figure}

In Fig.~\ref{fig:constraint_lepton}, we present analogous constraints and projections for DM annihilation channels that produce leptonic final states and generate neutrino and $\gamma$-ray signals. We derive Super-K and Hyper-K constraints and projections for the $\tau^{\pm}$ channel below 4~GeV, as well as Fermi-LAT constraints for the $\phi\to\mu^{\pm}$ channel and compare our result with previous Super-K limits from Refs.~\cite{Kappl:2011kz, Bernal:2012qh}. In the case of Ref.~\cite{Bernal:2012qh}, we note that our scattering constraints are based on different energy-cuts for the neutrino signal. While we consider the unattenuated neutrino signal above 100~MeV, Ref.~\cite{Bernal:2012qh} focuses on $\sim$20--60~MeV neutrinos produced through the cooling of muons and pions that are directly produced in the DM annihilation event. For the $\chi\chi\to2\phi\to4e$ annihilation channel, we compare our results with existing limits from Galileo Probe P2 observations of trapped $e^{\pm}$ in the Jovian magnetic field~\cite{Li:2022wix, Yan:2023kdg, Singh:2024nou}. For both scenarios we choose to use the same benchmark parameter $m_{\phi}> 2m_{\mu}$ for both $\phi\to e^{\pm}, \mu^{\pm}$ channels.

%\thong{Notably, all of our constraints exhibit a hard cutoff, where the limits become horizontal. This behavior arises because, even at the maximum geometric capture rate, the resulting neutrino and $\gamma$-ray fluxes remain too small to produce an observable signal. While the capture rate saturates at the geometric limit in Eq.~\ref{eq:Ctotal} as the cross section increases, the evaporation rate continues to grow, driving the DM population toward faster depletion and further suppressing the total annihilation signal.}

Compared with direct detection limits, our solar neutrino constraints from Super-K for the $\nu\bar{\nu}$ channel improve on existing direct detection bounds by 2--5 orders of magnitude in the DM mass range 2--4~GeV, and improve previous Super-K limits by more than one order of magnitude~\cite{Kappl:2011kz}. For the $\tau^{\pm}$ channel, the sensitivity reaches cross sections that are 2--4 orders of magnitude stronger than current direct detection limits and an order of magnitude stronger than previous Super-K limits. Corresponding Hyper-K projections further improve these constraints by a factor of approximately three relative to Super-K. Moreover, for the $\nu\bar{\nu}$ annihilation channel, our Hyper-K projections exceed previous Jupiter-based limits by roughly six orders of magnitude and approach the neutrino fog, where neutrino scattering with fluorine in direct detection experiments becomes a limiting background~\cite{OHare:2021utq}. Because the sensitivity of neutrino observations scales with exposure time, our approach is expected to probe this regime in future experiments.

For DM models that annihilate to long-lived mediators, our Fermi-LAT constraints probe spin-dependent cross sections down to $10^{-44}$~cm$^{2}$ for the $4\gamma$ channel, $10^{-42}$~cm$^{2}$ for the $4e$ channel, and $10^{-41}$~cm$^{2}$ for channels that directly annihilate to muons. These bounds improve upon current direct detection limits by 4--7 orders of magnitude and extend toward the neutrino fog regime. Moreover, these channels remain sensitive to DM masses below 2.3~GeV, where a two-order-of-magnitude gap exists between the PICASSO and CRESST limits. For the $4\gamma$ channel, solar constraints are approximately four orders of magnitude stronger than those derived from Jupiter observations in Ref.~\cite{Leane:2021tjj} for DM masses between 2 and 4~GeV. For the $4e$ channel, our Fermi-LAT $\gamma$-ray limits exceed those from Galileo Probe P2 observations of trapped $e^{\pm}$ in the Jovian magnetic field by about six orders of magnitude~\cite{Li:2022wix, Yan:2023kdg, Singh:2024nou}.

Notably, our constraints can even stretch down to 0.1~GeV, exploiting the very low evaporation masses that appear at very high cross sections in the LTE regime. We note that the choice to cut the analysis at 0.1~GeV matches atmospheric neutrino data from the Super-K collaboration~\cite{Super-Kamiokande:2015qek}. Our limits behave similarly to Super-K constraints from Ref.~\cite{Bernal:2012qh}, which take into account the full evaporation suppression of the LTE regime. Therefore, we show that for some highly motivated annihilation channels, such as $\chi\chi\to\nu\bar{\nu}$, existing Super-K observations can provide the strongest DM scattering constraint from 0.1--0.2~GeV, exceeding current limits from cosmic-ray boosted DM~\cite{Cappiello:2018hsu, Cappiello:2024acu}.

\section{Conclusion}
\label{sect:conclusion}
In this paper, we provide a detailed study of current and near future neutrino constraints, as well as the first $\gamma$-ray constraint on the spin-dependent DM capture cross section for DM masses near the putative ``evaporation limit" of solar DM searches. Building upon detailed calculations of evaporation by Refs.~\cite{Garani:2017jcj, Busoni:2017mhe, Leane:2022hkk, Bernal:2012qh}, we show that the evaporation mass is not a strict cutoff for the sensitivity of current instruments, and that existing solar observations can provide world leading sensitivity to spin-dependent DM scattering for DM masses that are as low as $\sim$2.6~GeV for direct annihilation into neutrinos, and down to $\sim$2~GeV for the annihilation into $\gamma$-rays through a light mediator. These constraints significantly extend the 4~GeV cutoff typically used in observational studies. While the cross-section limits for spin-dependent DM become significantly weaker at low masses, they do not disappear entirely. Intriguingly, we find our constraints again exceed terrestrial experiments in the mass range between 0.1--0.2~GeV for annihilation through both channels.

To reach these limits, we compute the accurate neutrino flux from DM annihilation into $\nu\bar{\nu}$, $\mu^+\mu^-$, $\tau^+\tau^-$, and $e^+e^-$ final states, as well as $\gamma$-ray constraints from annihilation through a pair of light mediators that then decay into $\mu^+\mu^-$, $e^+e^-$ and $\gamma\gamma$. Comparing these fluxes against current observations by Super-K (neutrinos) and the Fermi-LAT ($\gamma$-rays) we find limits that exceed leading direct detection constraints by up to two orders of magnitude, and previous Super-K constraints by more than one order of magnitude. For models where DM annihilates through a long-lived mediator, we obtain $\gamma$-ray constraints that exceed terrestrial observations by five orders of magnitude. We also predict that future Hyper-K observations will improve neutrino constraints by approximately a factor of three.

These results highlight the unique advantage of the Sun as a probe of DM interactions in celestial bodies, and motivate future searches for DM annihilation in astrophysical objects that push beyond the evaporation limit~\cite{Acevedo:2023owd}, explore additional targets~\cite{Blanco:2023qgi, Blanco:2024lqw, Linden:2025xom, Cappiello:2025yfe}, and develop complementary observational strategies~\cite{Feng:2016ijc, Feng:2015hja, Chauhan:2023zuf, Acevedo:2024ttq, Kouvaris:2016ltf, Braat:2025jvg, Linden:2024uph, Nguyen:2025eva, delaTorre:2023nfk, Chu:2024gpe}.

\vspace{0.5cm}
\section*{Acknowledgement}
We thank Chris Cappiello and Rebecca Leane for their comments on the manuscript, and Ciaran O'Hare for maintaining a detailed github repository on current DM cross section constraints~\href{https://github.com/cajohare/DirectDetectionPlots}{\faGithub}. We are particularly grateful to Sergio Palomares-Ruiz for critical comments that improved the mansucript and significantly improved the extension of our constraints to the sub-GeV regime. TTQN thanks the Luxembourg and Stockholm airports for wonderful internet connection and awesome working space where a large portion of this work was conducted. TTQN and TL are supported by the Swedish Research Council under contract 2022-04283. TL is also supported by the Swedish National Space Agency under contract 2023-00242. TL also acknowledges sabbatical support from
the Wenner-Gren foundation under contract SSh2024-0037. Parts of this work were performed using computing resources provided by the National Academic Infrastructure for Supercomputing in Sweden~(NAISS) under Projects~\mbox{2024/5-666} and~\mbox{2024/6-339}, which is partially funded by the Swedish Research Council through Grant~\mbox{2022-06725}.

\bibliography{ref}

\end{document}